\documentclass{llncs}

\usepackage{graphicx}
\graphicspath{{figures/}}
\usepackage{subcaption}
\usepackage{tikz}
\usetikzlibrary{decorations.pathmorphing}
\usetikzlibrary{calc}
\tikzset{full/.style={circle,fill,inner sep=0.05cm}}
\tikzset{empty/.style={circle,draw,inner sep=0.05cm}}
\tikzset{every path/.style={thick}}
\usepackage{paralist}
\usepackage{amsmath,dsfont}
\usepackage{hyperref}

\usepackage[nameinlink,capitalise]{cleveref}
\usepackage[utf8]{inputenc}
\hypersetup{colorlinks,linkcolor={blue},citecolor={blue},urlcolor={blue}}

\usepackage{cleveref}
\Crefname{observation}{Observation}{Observations}
\Crefname{algorithm}{Algorithm}{Algorithms}
\Crefname{section}{Sect.}{Sects.}
\Crefname{observation}{Observation}{Observations}
\Crefname{lemma}{Lemma}{Lemmas}
\Crefname{claim}{Claim}{Claims}
\Crefname{figure}{Fig.}{Figs.}
\Crefname{figure}{Fig.}{Figs.}
\Crefname{enumi}{Property}{Properties}
\Crefname{property}{Property}{Properties}
\Crefname{remark}{Remark}{Remarks}

\usepackage{color}

\newcommand{\PCOD}{PCOD}
\newcommand{\PCODs}{PCODs}
\newcommand{\complexity}{split complexity}
\newcommand{\Complexity}{Split complexity}
\newcommand{\SComplexity}{Split Complexity}

\renewcommand{\emph}[1]{{\color{blue}\em {#1}}}

\title{Planar Confluent Orthogonal Drawings of 4-Modal Digraphs}

\titlerunning{Planar Confluent Orthogonal Drawings of 4-Modal Digraphs}

\author{%
    Sabine Cornelsen \orcidID{0000-0002-1688-394X}%
    \thanks{The work of Sabine~Cornelsen was funded by the
    German Research Foundation DFG – Project-ID 50974019 – TRR 161 (B06).}
    \and
    Gregor Diatzko \orcidID{0000-0002-0904-4910}
}

\authorrunning{Cornelsen and Diatzko}

\institute{%
    University of Konstanz, Germany \\
    \href{mailto:sabine.cornelsen@uni-konstanz.de}{sabine.cornelsen@uni-konstanz.de},
    \href{mailto:gregor.diatzko@uni-konstanz.de}{gregor.diatzko@uni-konstanz.de}
}

\begin{document}
\color{black}

\maketitle

%%%%%%%%%%%%%%%%%%%%%%%%%%%%%%%%%%%%%%%%%%%%%%%
%
%%%%%%%%%%%%%%%%%%%%%%%%%%%%%%%%%%%%%%%%%%%%%%%

\begin{abstract}
    In a \emph{planar confluent orthogonal drawing (\PCOD)} of a directed graph
    (digraph) vertices are drawn as points in the plane and edges as orthogonal
    polylines starting with a vertical segment and ending with a horizontal
    segment.
    Edges may overlap in their first or last segment, but must not intersect
    otherwise. \PCODs{} can be seen as a directed variant of Kandinsky drawings
    or as planar L-drawings of subdivisions of digraphs.
    The maximum number of subdivision vertices in an edge is then the
    \emph{\complexity}.
    A \PCOD{} is \emph{upward} if each edge is drawn with monotonically
    increasing y-coordinates and \emph{quasi-upward} if no edge starts with
    decreasing y-coordinates.
    We study the \complexity{} of \PCODs{} and (quasi-)upward \PCODs{} for
    various classes of graphs.

    \keywords{%
        directed plane graphs,
        Kandinsky drawings,
        L-drawings,
        curve complexity,
        irreducible triangulations,
        (quasi-)upward planar
    }
\end{abstract}

%%%%%%%%%%%%%%%%%%%%%%%%%%%%%%%%%%%%%%%%%%%%%%%
%
%%%%%%%%%%%%%%%%%%%%%%%%%%%%%%%%%%%%%%%%%%%%%%%

\section{Introduction}

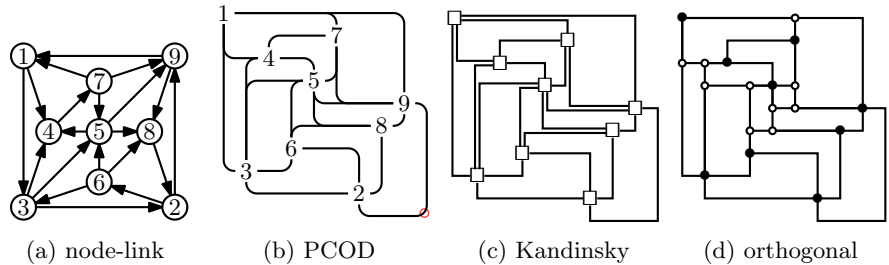
\begin{figure}[t]
    \centering
    \subcaptionbox{node-link\label{SUBFIG:graph}}{
        \includegraphics[page=2]{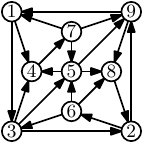}%
    }\hfil%
    \subcaptionbox{PCOD\label{SUBFIG:introPCOD}}{
        \begin{tikzpicture}[scale=.3]%{{{
            \tikzset{every path/.style={thick,rounded corners}}
            \tikzset{every node/.style={circle,inner sep=0.04}}
            \node (1) at (1,10) {$1$};
            \node (2) at (7,2) {$2$};
            \node (3) at (2,3) {$3$};
            \node (4) at (3,8) {$4$};
            \node (5) at (5,7) {$5$};
            \node (6) at (4,4) {$6$};
            \node (7) at (6,9) {$7$};
            \node (8) at (8,5) {$8$};
            \node (9) at (9,6) {$9$};
            \node[red] (29) at (9.9,1.1) {$\circ$};

            \draw (7) |- (1);
            \draw (9) |- (1);
            \draw (1) |- (4);
            \draw (1) |- (3);
            \draw (6) |- (5);
            \draw (7) |- (9);
            \draw (7) |- (5);
            \draw (4) |- (7);
            \draw (3) |- (4);
            \draw (3) |- (5);
            \draw (6) |- (3);
            \draw (3) |- (2);
            \draw (2) |- (10,1) |- (9);
            \draw (2) |- (6);
            \draw (9) |- (8);
            \draw (5) |- (9);
            \draw (5) |- (4);
            \draw (5) |- (8);
            \draw (6) |- (8);
            \draw (8) |- (2);
        \end{tikzpicture}%}}}
    }\hfil%
    \subcaptionbox{Kandinsky\label{SUBFIG:introKandinsky}}{%
        \begin{tikzpicture}[scale=.3]%{{{
            \tikzset{>=latex}
            \tikzset{every node/.style={circle,inner sep=-1.2}}
            \node (1) at (1,10){$\qed$};%{$1$};
            \node (2) at (7,2) {$\qed$};%{$2$};
            \node (3) at (2,3) {$\qed$};%{$3$};
            \node (4) at (3,8) {$\qed$};%{$4$};
            \node (5) at (5,7) {$\qed$};%{$5$};
            \node (6) at (4,4) {$\qed$};%{$6$};
            \node (7) at (6,9) {$\qed$};%{$7$};
            \node (8) at (8,5) {$\qed$};%{$8$};
            \node (9) at (9,6) {$\qed$};%{$9$};

            \draw (7) |- ($ (1) + ( .29,-0.1 ) $);
            \draw (9) |- ($ (1) + ( .29,+0.1 ) $);
            \draw ($ (1) + ( +0.1,-.29 ) $) |- ($ (4) + ( -.29,+0.1 ) $);
            \draw ($ (1) + ( -0.1,-.29 ) $) |- (3);
            \draw ($ (6) + ( -0.1,+.29 ) $) |- ($ (5) + ( -0.29,-0.1 ) $);
            \draw ($ (7) + ( +0.1,-.29 ) $) |- ($ (9) + ( -0.29,+0.1 ) $);
            \draw ($ (7) + ( -0.1,-0.29 ) $) |- ($ (5) + ( +0.29,0 ) $);
            \draw (4) |-  (7);
            \draw ($ (3) + ( -0.1,+.29 ) $) |- ($ (4) + ( -0.29,-0.1 ) $);
            \draw ($ (3) + ( +0.1,+.29 ) $) |- ($ (5) + ( -0.29,+0.1 ) $);
            \draw (6) |-  (3);
            \draw (3) |- (2);
            \draw (2) |- (10,1) |- (9);
            \draw (2) |- (6);
            \draw (9) |- (8);
            \draw ($ (5) + ( +0.1,-.29 ) $) |- ($ (9) + ( -0.29,-0.1 ) $);
            \draw (5) |- (4);
            \draw ($ (5) + ( -0.1,-.29 ) $) |- ($ (8) + ( -0.29,+0.1 ) $);
            \draw ($ (6) + ( +0.1,+.29 ) $) |- ($ (8) + ( -0.29,-0.1 ) $);
            \draw (8) |- (2);
        \end{tikzpicture}%}}}
    }\hfil%
    \subcaptionbox{orthogonal\label{SUBFIG:introOrthogonal}}{%
        \begin{tikzpicture}[scale=.3]%{{{
            \tikzset{every node/.style={circle,inner sep=-2}}
            \node (1) at (1,10) {$\bullet$};
            \node (2) at (7,2) {$\bullet$};
            \node (3) at (2,3) {$\bullet$};
            \node (4) at (3,8) {$\bullet$};
            \node (5) at (5,7) {$\bullet$};
            \node (6) at (4,4) {$\bullet$};
            \node (7) at (6,9) {$\bullet$};
            \node (8) at (8,5) {$\bullet$};
            \node (9) at (9,6) {$\bullet$};

            \draw (7) |- (1);
            \draw (9) |- (1);
            \draw (1) |- (4);
            \draw (1) |- (3);
            \draw (6) |- (5);
            \draw (7) |- (9);
            \draw (7) |- (5);
            \draw (4) |- (7);
            \draw (3) |- (4);
            \draw (3) |- (5);
            \draw (6) |- (3);
            \draw (3) |- (2);
            \draw (2) |- (10,1) |- (9);
            \draw (2) |- (6);
            \draw (9) |- (8);
            \draw (5) |- (9);
            \draw (5) |- (4);
            \draw (5) |- (8);
            \draw (6) |- (8);
            \draw (8) |- (2);

            \node (14) at (1,8) {};
            \node (71) at (6,10) {};
            \node (35) at (2,7) {};
            \node (34) at (2,8) {};
            \node (79) at (6,6) {};
            \node (75) at (6,7) {};
            \node (65) at (4,7) {};
            \node (68) at (4,5) {};
            \node (59) at (5,6) {};
            \node (58) at (5,5) {};

            \draw[fill=white] (14) circle (0.15);
            \draw[fill=white] (71) circle (0.15);
            \draw[fill=white] (35) circle (0.15);
            \draw[fill=white] (34) circle (0.15);
            \draw[fill=white] (79) circle (0.15);
            \draw[fill=white] (75) circle (0.15);
            \draw[fill=white] (65) circle (0.15);
            \draw[fill=white] (68) circle (0.15);
            \draw[fill=white] (59) circle (0.15);
            \draw[fill=white] (58) circle (0.15);
        \end{tikzpicture}%}}}
    }%
    \caption{\label{FIG:intro}Different representations of a 4-modal irreducible
    triangulation.}
\end{figure}

We consider \emph{plane digraphs}, i.e., planar directed graphs with a fixed
planar embedding and a fixed outer face. Directions of edges in node-link
diagrams are usually indicated by arrow heads. Since this might cause clutter at
vertices with high indegree, Angelini et al.~\cite{angeliniLdrawings} proposed
L-drawings in which each edge is drawn with a 1-bend orthogonal polyline
starting with a vertical segment at the tail. A plane digraph can only have an
L-drawing without crossings if it is \emph{4-modal}, where a plane digraph is
\emph{$k$-modal} if in the cyclic order around a vertex there are at most $k$
pairs of consecutive edges that are neither both incoming nor both outgoing.
However, not every 4-modal digraph admits a planar L-drawing. This motivates to
extend the model to drawings with more than one bend per edge.

In a \emph{planar confluent orthogonal drawing (\PCOD)}  of a digraph, vertices
are represented as points in the plane with distinct x- and y-coordinates and
each edge is represented as an orthogonal  polyline starting with a vertical
segment at the tail and ending with a horizontal segment at the head.
Distinct edges may overlap in a first or last segment, but must not intersect
otherwise.
For better readability bends have distinct coordinates and are drawn with
rounded corners.
A plane digraph has a \PCOD{} if and only if it is 4-modal.

A \PCOD{} of a digraph $G$ corresponds to a planar L-drawing of a subdivision of
$G$. The number of subdivision vertices on an edge is its \emph{\complexity{}}.
See the red encircled vertex in \cref{SUBFIG:introPCOD} for the subdivision
vertex.
Since each edge starts with a vertical segment and ends with a horizontal
segment, the number of bends on an edge is odd. An edge with \complexity{} $k$
has $2k+1$ bends. The \emph{\complexity{}} of a \PCOD{} is the maximum
\complexity{} of any edge. The \PCOD{} in \cref{SUBFIG:introPCOD} has
\complexity{} one.
A \emph{planar L-drawing}~\cite{angeliniLdrawings,chaplick_etal:gd17} is a
\PCOD{} of \complexity{} zero.
If the embedding is not fixed, then it is NP-complete to decide whether a
digraph admits a planar L-drawing~\cite{chaplick_etal:gd17}.
Every 2-modal digraph without 2-cycles has a
planar~L-drawing~\cite{angelini_etal:jgaa22}.

A \PCOD{} of a digraph corresponds to a Kandinsky
drawing~\cite{foessmeier/kaufmann:gd95} of the underlying undirected graph with
the only difference that edges partially overlap instead of being drawn in
parallel with a small gap.
See \cref{SUBFIG:introKandinsky}.
While every simple planar graph has a Kandinsky drawing with one bend per
edge~\cite{brueckner:ba13}, deciding whether a multigraph has a Kandinsky
drawing with one bend per edge~\cite{brueckner:ba13} or finding the minimum
number of bends in a Kandinsky drawing of a plane
graph~\cite{blaesius_etal:esa14} is NP-hard.
For the bend-minimization problem in the Kandinsky model there are
2-approximization algorithms~\cite{eiglsperger:phd,barth_etal:gd06} and
heuristics~\cite{bekos_etal:sea15}.

Among the results for orthogonal drawings of undirected graphs where edges must
not overlap, we mention three:
With one exception, every plane graph of maximum degree four admits an
orthogonal drawing with at most two bends per edge~\cite{biedl/kant:98}.
In a bend-minimum drawing, however, there might have to be an edge with a linear
number of bends~\cite{tamassia/tollis/vitter:1991}.
An orthogonal drawing with the minimum number of bends can be computed by means
of a min-cost flow approach~\cite{tamassia:87} even if an upper bound on the
number of bends per edge must be respected.

A \PCOD{} is \emph{upward} if each edge is drawn with monotonically increasing
y-coordinates.
A digraph is \emph{upward-planar} if and only if it has an upward \PCOD{}.
A \emph{plane $st$-graph}, i.e., a plane acyclic digraph with a single sink and
a single source, both on the outer face, is always upward-planar; moreover, it
has an upward-planar L-drawing if and only if it admits a so-called bitonic
$st$-ordering~\cite{chaplick_etal:gd17}.
Since it suffices to subdivide the edges of a plane $st$-graph at most once in
order to obtain a digraph that admits a bitonic
$st$-ordering~\cite{gronemann:gd16,angelini_etal:wg20}, it follows that every
plane $st$-graph admits an upward \PCOD{} with \complexity{} one.
Moreover, the minimum number of bends in an upward \PCOD{} of a plane $st$-graph
can be determined in linear time.
In general, a digraph admits an upward-planar L-drawing, if and only if it is a
subgraph of a plane $st$-graph admitting a bitonic
$st$-ordering~\cite{angelini_etal:mfcs22}.
Not every 2-modal tree admits an upward-planar
L-drawing~\cite{angelini_etal:mfcs22}.

In a \emph{quasi-upward-planar drawing}~\cite{bertolazzi_etal:quasiUpward:2002}
edges must be strictly monotonically increasing in y-direction in a small
vicinity around the end vertices.
A digraph has a 2-modal embedding if and only if it admits a quasi-upward-planar
drawing.
Every 2-modal graph without 2-cycles admits a quasi-upward planar drawing with
at most two bends per edge and the curve complexity in such drawings can be
minimized utilizing a min-cost flow approach \cite{binucci_etal:gd21}.
We call a \PCOD{} \emph{quasi-upward} if no edge starts with decreasing
y-coordinates.

\paragraph{Our Contribution.}
We show that \PCODs{} of 4-modal trees have \complexity{} zero
(\cref{THEO:trees}), \complexity{} two is sufficient (\cref{THEO:multi}) and
sometimes necessary (\cref{THEO:multi-examples}) for \PCODs{} of 4-modal
digraphs with parallel edges or loops, while \complexity{} one suffices for
4-modal \emph{irreducible triangulations} (\cref{THEO:irred}), i.e., internally
triangulated 4-connected graphs with an outer face of degree 4.
\Complexity{} one also suffices for upward \PCODs{} of  upward-plane digraphs
(\cref{THEO:upward}) and for quasi-upward \PCODs{} of 2-modal digraphs without
2-cycles (\cref{THEO:quas-upward}).
Using an ILP, we conducted experiments that suggest that every simple 4-modal
digraph without separating 2-cycles admits a \PCOD{}  with \complexity{} one
(\cref{SEC:ilp}).
Constant \complexity{} is not to be expected for bend-minimum \PCODs{}
(\cref{THEO:linearComplexity}).

%%%%%%%%%%%%%%%%%%%%%%%%%%%%%%%%%%%%%%%%%%%%%%%
%
%%%%%%%%%%%%%%%%%%%%%%%%%%%%%%%%%%%%%%%%%%%%%%%

\section{Preliminaries}\label{SEC:prel}

Two consecutive incident edges of a vertex $v$ are a \emph{switch} if both edges
are incoming or both outgoing edges of $v$.
The drawing of a \PCOD{} is determined by the coordinates of the vertices and
the coordinates of every second bend of an edge.
We call a bend \emph{independent} if it is the second, fourth, etc.\  bend of an
edge.
Considering a \PCOD{} as an L-drawing of a subdivision, the independent bends
correspond to the subdivision vertices.
The \complexity{} of an edge is the number of its independent bends.
The total number of bends equals the number of edges plus twice the number of
independent bends.
The top, left, bottom, and right side of a vertex is its \emph{North, West,
South}, and \emph{East port}, respectively.

An \emph{$st$-ordering} of a biconnected (undirected) graph $G=(V,E)$ is a
bijection $\pi: V \rightarrow \{1,\dots,|V|\}$ such that $\pi(s) = 1$, $\pi(t) =
|V|$, and each vertex $v \in V \setminus \{s,t\}$ has neighbors $u$ and $w$ with
$\pi(u) < \pi(v) < \pi(w)$.
Let now $G=(V,E)$ be a plane $st$-graph.
If $(v,v_i)$, $i=1,\dots,k$ are the outgoing edges of a vertex $v$ from left to
right then $S(v)=\left<v_1,\dots,v_k\right>$ is the \emph{successor list} of
$v$.
A \emph{bitonic $st$-ordering} of $G$ is a bijection $\pi: V \rightarrow
\{1,\dots,|V|\}$ such that $\pi(u) < \pi(v)$ for $(u,v) \in E$ and
$S(v)=\left<v_1,\dots,v_k\right>$ is \emph{bitonic} for each vertex $v$, i.e.,
there is a $1 \leq h \leq k$ such that $\pi(v_i)<\pi(v_{i+1})$, $i=1,\dots,h-1$
and $\pi(v_i)>\pi(v_{i+1})$, $i=h,\dots,k-1$.
The successor list $S(v)=\left<v_1,\dots,v_k\right>$ contains a \emph{valley}
with transitive edges $(v,v_{\ell-1})$ and $(v,v_{r+1})$ if there is a directed
$v_\ell$-$v_{\ell-1}$-path and a directed $v_r$-$v_{r+1}$-path for some $1 <
\ell \leq r < k$.
A plane $st$-graph admits a bitonic $st$-ordering if and only if it does not
contain a valley \cite{gronemann:gd16}.

%%%%%%%%%%%%%%%%%%%%%%%%%%%%%%%%%%%%%%%%%%%%%%%
%
%%%%%%%%%%%%%%%%%%%%%%%%%%%%%%%%%%%%%%%%%%%%%%%

\section{Confluent Orthogonal Representation}\label{SEC:representation}

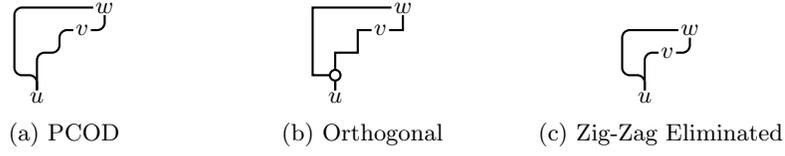
\begin{figure}[t]
    \centering
    \subcaptionbox{PCOD\label{SUBFIG:zigzagPCOD}}[0.3\textwidth]{%
        \begin{tikzpicture}[scale=.3]%{{{
            \tikzset{every path/.style={thick,rounded corners=1mm}}
            \tikzset{every node/.style={inner sep=.5pt}}
            \node (u) at (1,0) {$u$};
            \node (v) at (3,3) {$v$};
            \node (w) at (4,4) {$w$};

            \draw (u) |- (2,2) |- (v);
            \draw (u) |- (0,1) |- (w);
            \draw (w) |- (v);
        \end{tikzpicture}%}}}
    }\hfil%
    \subcaptionbox{Orthogonal\label{SUBFIG:zigzagOrthogonal}}[0.3\textwidth]{%
        \begin{tikzpicture}[scale=.3]%{{{
            \tikzset{every path/.style={thick}}
            \tikzset{every node/.style={inner sep=.5pt}}
            \node (u) at (1,0) {$u$};
            \node (v) at (3,3) {$v$};
            \node (w) at (4,4) {$w$};

            \draw (u) |- (2,2) |- (v);
            \draw (u) |- (0,1) |- (w);
            \draw (w) |- (v);

            \node[empty,fill=white] (x) at (1,1) {};
        \end{tikzpicture}%}}}
    }\hfil%
    \subcaptionbox{Zig-Zag Eliminated\label{SUBFIG:zigzagEliminated}}[0.3\textwidth]{%
        \begin{tikzpicture}[scale=.3]%{{{
            \tikzset{every path/.style={thick,rounded corners=1mm}}
            \tikzset{every node/.style={inner sep=.5pt}}
            \node (u) at (1,0) {$u$};
            \node (v) at (2,2) {$v$};
            \node (w) at (3,3) {$w$};

            \draw (u) |- (v);
            \draw (u) |- (0,1) |- (w);
            \draw (w) |- (v);
        \end{tikzpicture}%}}}
    }%
    \caption{\label{FIG:zigzag}Eliminating zig-zags.}
\end{figure}

Let $\Gamma$ be a \PCOD{} of a plane digraph $G$.
We call a bend \emph{covered} if it is contained in the drawing of another edge.
We associate an orthogonal drawing of a plane graph $G_\Gamma$ with $\Gamma$ as
follows \cite{angelini_etal:jgaa22}:
Replace every covered bend in $\Gamma$ by a dummy vertex.
See \cref{SUBFIG:introOrthogonal,SUBFIG:zigzagOrthogonal}.
A \emph{zig-zag} is a pair of  uncovered bends on an edge, one with a left turn,
and one with a right turn.
E.g., on the edge $(u,v)$ in \cref{SUBFIG:zigzagPCOD} there is a zig-zag, while
on the edge $(u,w)$ there is both a left and a right turn, but the left turn is
covered, so there is no zig-zag.
Since the number of bends in an orthogonal drawing can always be reduced by
eliminating zig-zags, we will also do so in \PCODs{} (see
\cref{SUBFIG:zigzagEliminated}) and, thus, the ordering of left- and right-turns
at uncovered bends of an edge will not matter.
Since planar (confluent) orthogonal drawings can be stretched independently in
x- and y-directions, it is algorithmically often easier not to work with actual
x- and y-coordinates, but rather with the shape of the faces in terms of bends
on the edges and angles at the vertices.
See also \cite{tamassia:87,foessmeier/kaufmann:gd95}.

A \emph{confluent orthogonal representation} $R$ of a plane digraph $G = (V,E)$
is a set of circular lists $H(f)$, one for each face $f$ of $G$.
The elements of $H(f)$ are tuples $r = (e,v,a,s,b)$ associated with the edges
$e$ incident to $f$ in counter-clockwise order.
\begin{inparaenum}[(a)]
    \item
        $v$ is the end vertex of $e$ traversed immediately before~$e$.
    \item
        $a \in \{0,\frac\pi2,\pi,\frac{3\pi}2,2\pi\}$ is the angle at $v$
        between $e$ and its predecessor on $f$.
        It is a multiple of $\pi$ if and only if it describes an angle at a
        switch.
    \item
        $s$ is the number of left turns (when traversing $e$ starting from $v$)
        at bends in $e$.
    \item
        $b \in \{L,N,R\}$ represents a covered bend on the segment of $e$
        incident to $v$, if any, with ($L$) a left bend, ($R$) a right bend, or
        ($N$) no such bend.
\end{inparaenum}

Let $r_p$ be the predecessor of $r$ in $H(f)$.
If $r$ is not clear from the context, we denote the entries by
$e[r],v[r],a[r],s[r],b[r]$.
Each edge is contained twice in a confluent orthogonal representation.
Let $\overline{r}$ be the (other) entry containing $e[r]$.
A confluent orthogonal representation is \emph{feasible} if it fulfills the
following.

\begin{inparaenum}[(i)]
    \item\label{PROP:rotation}
        The rotation $\sum_{r \in H(f)} (2 - a[r] / \frac\pi2 + s[r] -
        s[\overline r])$ of a face $f$  is  $-4$ if $f$ is the outer face and
        $4$ otherwise.
    \item\label{PROP:angularSum}
        The angular sum $\sum_{r; v[r]=v} a[r]$ around a vertex $v$ is $2\pi$.
        \item\label{PROP:coveredBends}
            If $b[r] = L$ or $b[\overline r] = R$ then $s[r] \geq 1$ and if both
            $b[r] = L$ and $b[\overline r] = R$ then $s[r] \geq 2$.
            This ensures that covered  bends are counted by $s$ and that covered
            bends adjacent to the head or the tail of an edge must be distinct.
    \item\label{PROP:bendorend}
        The so called \emph{bend-or-end property}, i.e., if $a[r] = 0$, then
        $b[r] = R$ or $b[\overline {r_p}] = L$.
    \item\label{PROP:oddbends}
        The total number of bends $s[r] + s[\overline r]$ on $e[r]$ is odd.
\end{inparaenum}

\paragraph{From a Representation to a \PCOD.}
In order to construct a \PCOD{} from a feasible confluent orthogonal
representation $R$ of a plane digraph $G$, we transform $G$ into a graph $G_R$
of  maximum degree $4$ and a feasible orthogonal representation $R'$ without $0$
or $2\pi$ angles.
Using compaction for orthogonal representations \cite{tamassia:87} on $G_R$ then
yields a  \PCOD{} or a $\pi/2$-rotation of a \PCOD{} in linear time.
The idea for the construction of $G_R$ is analogous to the construction of
$G_\Gamma$ from a \PCOD{} $\Gamma$ and is as follows:
Consider a vertex $v \in V$ and let $e_1,\dots,e_k$ be a maximum sequence of
consecutive edges around $v$ with 0 angles.
Let $r_i$, $i=1,\dots,k$ be the entry with $e[r_i]=e_i$ and $v[r_i]=v$.
Due to \cref{PROP:bendorend}, there is a $1 \leq m \leq k$ such that $b[r_j]=L$,
$j < m$ and $b[r_j]=R$, $m < j$.
We subdivide the segment of $e_m$ that is incident to $v$ with $k - 1$ vertices
$v_1,\dots,v_{m-1},v_k,\dots,v_{m+1}$ in this order, starting from $v$.
We attach $e_j,j \not= m$ to $v_j$ instead of $v$.
The representation $R'$ is updated accordingly.

%%%%%%%%%%%%%%%%%%%%%%%%%%%%%%%%%%%%%%%%%%%%%%%
%
%%%%%%%%%%%%%%%%%%%%%%%%%%%%%%%%%%%%%%%%%%%%%%%

\section{Some Initial Results}\label{SEC:firstResults}

\begin{figure}[t]
    \center
    \subcaptionbox{$G_k$\label{SUBFIG:Gk}}{%
        \begin{tikzpicture}[scale=0.8]%{{{
            \tikzset{leaf/.style={circle,fill,inner sep=0.04cm}}
            \tikzset{>=latex}

            \node (0)  at (2,2) {$s_1$};
            \node (1)  at (2,4) {$t_1$};
            \node (2)  at (2,6) {$s_2$};
            \node (p)  at (2,7) {$s_k$};
            \node (u)  at (2,9) {$t_k$};
            \node (01) at (2,3) {$x_1$};
            \node (12) at (2,5) {$y_1$};
            \node (pu) at (2,8) {$x_k$};
            \node (dots) at (2,6.6) {$\vdots$};

            \draw[->] (0) to[bend right] (1);
            \draw[<-] (1) to[bend right] (2);
            \draw[->] (p) to[bend right] (u);
            \draw[->] (0) to (01);
            \draw[->] (01) to (1);
            \draw[<-] (1) to (12);
            \draw[<-] (12) to (2);
            \draw[->] (p) to (pu);
            \draw[->] (pu) to (u);

            \foreach \v in {0,2,p}{
                \node[leaf] (1l\v) at ($ (\v) + (200:1) $) {};
                \node[leaf] (2l\v) at ($ (\v) + (180:1) $) {};
                \node[leaf] (3l\v) at ($ (\v) + (160:1) $) {};
                \draw[->] (1l\v) to (\v);
                \draw[<-] (2l\v) to (\v);
                \draw[->] (3l\v) to (\v);
            }
            \foreach \v in {1,u}{
                \node[leaf] (1l\v) at ($ (\v) + (200:1) $) {};
                \node[leaf] (2l\v) at ($ (\v) + (180:1) $) {};
                \node[leaf] (3l\v) at ($ (\v) + (160:1) $) {};
                \draw[<-] (1l\v) to (\v);
                \draw[->] (2l\v) to (\v);
                \draw[<-] (3l\v) to (\v);
            }
            \foreach \v in {01,pu}{
                \node[leaf] (1l\v) at ($ (\v) + (170:1) $) {};
                \node[leaf] (2l\v) at ($ (\v) + (190:1) $) {};
                \draw[->] (1l\v) to (\v);
                \draw[<-] (2l\v) to (\v);
            }
            \node[leaf] (1l12) at ($ (12) + (170:1) $) {};
            \node[leaf] (2l12) at ($ (12) + (190:1) $) {};
            \draw[<-] (1l12) to (12);
            \draw[->] (2l12) to (12);

            \draw[->,red,dashed] (0.south)
                arc (0:-180:1)
                to ($ (u.north) - (2,0) $)
                arc (180:0:1);
        \end{tikzpicture}%}}}
    }
    \hfil%
    \subcaptionbox{Bend-Minimum \PCOD{} of $G_2$\label{SUBFIG:G2}}{%
        \begin{tikzpicture}[scale=.3]%{{{
            \tikzset{every path/.style={thick,rounded corners=2}}
            \tikzset{leaf/.style={circle,fill,inner sep=0.04cm}}

            \node       (0)  at ( 7, 5) {$s_1$};
            \node       (1)  at ( 5, 8) {$x_1$};
            \node       (2)  at ( 3,10) {$t_1$};
            \node       (3)  at (11,12) {$y_1$};
            \node       (4)  at (13,14) {$s_2$};
            \node       (5)  at (15, 1) {$x_2$};
            \node       (6)  at (17,-1) {$t_2$};
            \node[leaf] (l1) at (6,5.5) {};
            \node[leaf] (l2) at (8,4.5) {};
            \node[leaf] (l3) at (6.5,4) {};
            \node[leaf] (l4) at (4,8.5) {};
            \node[leaf] (l5) at (4.5,7) {};
            \node[leaf] (l6) at (2,10.5) {};
            \node[leaf] (l7) at (2.5,9) {};
            \node[leaf] (l8) at (3.5,11) {};
            \node[leaf] (l9) at (10,12.5) {};
            \node[leaf] (l10) at (11.5,13) {};
            \node[leaf] (l11) at (12,14.5) {};
            \node[leaf] (l12) at (13.5,15) {};
            \node[leaf] (l13) at (14,13.5) {};
            \node[leaf] (l14) at (15.5,2) {};
            \node[leaf] (l15) at (16,0.5) {};
            \node[leaf] (l16) at (17.5,0) {};
            \node[leaf] (l17) at (18,-1.5) {};
            \node[leaf] (l18) at (16.5,-2) {};

            \draw (0) -| (l1);
            \draw (0) -| (l2);
            \draw (0) |- (l3);
            \draw (0) |- (1);
            \draw (0) |- (2);
            \draw (1) -| (l4);
            \draw (1) |- (l5);
            \draw (1) |- (2);
            \draw (2) -| (l6);
            \draw (2) |- (l7);
            \draw (2) |- (l8);
            \draw (3) |- (2);
            \draw (3) -| (l9);
            \draw (3) |- (l10);
            \draw (4) |- (2);
            \draw (4) |- (3);
            \draw (4) -| (l11);
            \draw (4) |- (l12);
            \draw (4) -| (l13);
            \draw (4) |- (5);
            \draw (4) |- (6);
            \draw (5) |- (l14);
            \draw (5) -| (l15);
            \draw (5) |- (6);
            \draw (6) |- (l16);
            \draw (6) -| (l17);
            \draw (6) |- (l18);
            \draw[red,dashed] (0) |- (9,6) |- (1,3) |- (19,16) |- (-1,-3) |- (6);
        \end{tikzpicture}%}}}
    }
    \caption{Graphs with a linear number of bends in any bend-minimum \PCOD{}.}
    \label{FIG:gk}
\end{figure}
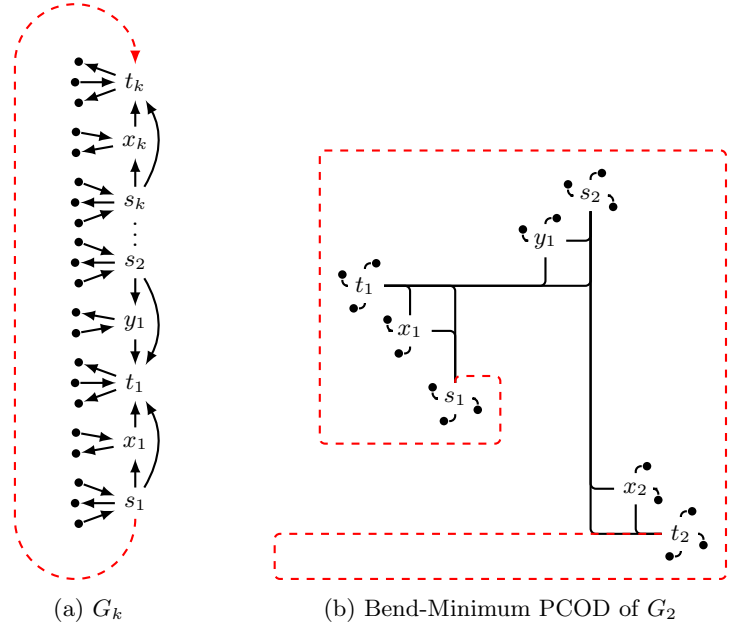

\begin{theorem}\label{THEO:linearComplexity}
    There is a family $G_k$, $k >0$ of 4-modal digraphs with $14k-3$ vertices
    and $16k-4$ edges such that in any bend-minimum \PCOD{} of $G_k$ there is an
    edge with \complexity{} at least $k+2$.
\end{theorem}
\begin{proof}
    Consider the digraphs $G_k$ indicated in \cref{SUBFIG:Gk}.
    Let $e$ be the red dashed edge.
    Let $P_k$ be the path $s_1,x_1,t_1,y_1,\dots,s_k,x_k,t_k$ of length $4k-2$
    in $G_k$ that is drawn vertically in  \cref{SUBFIG:Gk}.
    Consider a planar L-drawing of $G_k-e$ in which all edges of $P_k$
    (traversed from $s_1$ to $t_k$) bend to the left and the edge incident to
    $s_1$ is to the top of $s_1$.
    Such a drawing for $G_2$  is indicated in \cref{SUBFIG:G2}.
    Since all vertices of $P_k$ are 4-modal this uniquely determines the drawing
    of $P_k$ and also of the transitive edges of $P_k$.
    In order to preserve the embedding, $e$ can only be inserted into the
    drawing with \complexity{} at least $k+2$.

    Consider a \PCOD{} of $G_k$ with fewer bends on $e$.
    Since all vertices are 4-modal, the rotation of the cycle $C$ composed of
    $P_k$ and $e$ can only be maintained, if the number of bends on at least one
    edge of $P_k$, say $(s_i,x_i)$, is increased.
    But then we also must increment the number of bends on an edge $(s_i,t_i)$
    to maintain the rotation of the face bounded by the edges $(s_i,t_i),
    (s_i,x_i), (x_i,t_i)$.
    Thus, for each independent bend less on $e$ the total number of bends
    increases by at least 2.
    \qed
\end{proof}

Even though not every 2-modal tree has an upward-planar L-drawing
\cite{angelini_etal:mfcs22}, every 4-modal tree has a planar L-drawing, despite
its fixed embedding.

\begin{theorem}\label{THEO:trees}
    Every 4-modal tree has a  \PCOD{} with \complexity{} zero.
    Moreover, such a drawing can be constructed in linear time.
\end{theorem}
\begin{proof}
    Let $T$ be a 4-modal tree and let $v$ be a leaf of $T$.
    We show by induction on the number $m$ of edges that we can draw $T$ as a
    \PCOD{} $\Gamma^\alpha$  with \complexity{} zero such that $v$ is in the
    corner $\alpha$ (lower left ($\ell\ell$), lower right ($\ell r$), upper left
    ($u\ell$), upper right ($ur$)) of the bounding box of $\Gamma^\alpha$.
    We give the details for  $\Gamma^{\ell\ell}$; the other cases are analogous.
    If $m=1$, draw $v$ at $(0,0)$ and its neighbor at~$(1,1)$.

    If $m > 1$, let the neighbor of $v$ be $v'$, and let the connected
    components of \mbox{$T - v'$} be $v,T_1, \dots, T_k$ in clockwise order
    around $v'$.
    See \cref{FIG:tree}.
    Let $T_0$ be the subtree consisting of $v$ only.
    Each tree $T_i + v'$, $i = 0, \dots, k$ has at most $m - 1$ edges and the
    leaf $v'$; therefore, by the inductive hypothesis, we can construct \PCODs{}
    $\Gamma_i^\alpha$, $\alpha \in \{\ell\ell,\ell r, u\ell, ur\}$ of $T_i + v'$
    with $v'$ in the respective corner of the bounding box.
    W.l.o.g.\ let $v$ be the tail of the edge connecting $v$ and $v'$, see
    \cref{SUBFIG:gamma-lo}.
    Let $1 \leq a \leq b \leq c \leq d \leq k$ such that $T_{d+1}, \dots, T_k,
    T_0, T_1, \dots, T_a$ and $T_{b+1}, \dots, T_c$ are connected to $v'$ by an
    incoming edge and $T_{a+1}, \dots, T_b$ and $T_{c+1}, \dots, T_d$ by
    outgoing edges.
    Choose $\Gamma_0^{u r},\Gamma_1^{\ell r}, \dots, \Gamma_a^{\ell r},
    \Gamma_{a+1}^{\ell \ell}, \dots, \Gamma_c^{\ell \ell}, \Gamma_{c+1}^{ur},
    \dots,\Gamma_k^{ur}$ for $T_i+v'$, $i = 1, \dots, k$.
    Finally, merge the drawings of the subtrees at $v'$.

    \begin{figure}[t]
        \begin{minipage}{.65\textwidth}
            \hfil%
            \subcaptionbox{$\Gamma^{\ell\ell}$, edge $(v,v')$\label{SUBFIG:gamma-lo}}{%
                \begin{tikzpicture}[scale=.5]%{{{
                    \tikzset{every path/.style={thick,rounded corners}}

                    \node (l) at (0,0) {$v$};
                    \node (v) at (4,3) {$v'$};
                    \node[draw,fill=black!20] (t1) at (3,4) {$\Gamma_a^{\ell r}$};
                    \node[draw,fill=black!20] (t2) at (5,6) {$\Gamma_b^{\ell\ell}$};
                    \node[draw,fill=black!20] (t3) at (6.2,4.8) {$\Gamma_c^{\ell\ell}$};
                    \node[draw,fill=black!20] (t4) at (2.2,0.8) {${\Gamma_d^{ur}}$};
                    \node[draw,fill=black!20] (t5) at (1,2) {${\Gamma_k^{ur}}$};

                    \draw (v) -| (t1);
                    \draw (v) |- (t2);
                    \draw (v) -| (t3);
                    \draw (v) |- (t4);
                    \draw (v) -| (t5);
                    \draw (v) -| (l);
                \end{tikzpicture}%}}}
            }
            \hfil%
            \subcaptionbox{$\Gamma^{\ell\ell}$, edge $(v',v)$\label{SUBFIG:gamma-li}}{%
                \begin{tikzpicture}[scale=.5]%{{{
                    \tikzset{every path/.style={thick,rounded corners}}

                    \node (l) at (0,0) {$v$};
                    \node (v) at (4,3) {$v'$};
                    \node[draw,fill=black!20] (t1) at (3,4)     {$\Gamma_b^{\ell r}$};
                    \node[draw,fill=black!20] (t2) at (5,6)     {$\Gamma_c^{\ell\ell}$};
                    \node[draw,fill=black!20] (t3) at (6.2,4.8) {$\Gamma_d^{\ell\ell}$};
                    \node[draw,fill=black!20] (t4) at (2.2,0.8) {${\Gamma_a^{ur}}$};
                    \node[draw,fill=black!20] (t5) at (6,2)     {${\Gamma_k^{u\ell}}$};

                    \draw (v) -| (t1);
                    \draw (v) |- (t2);
                    \draw (v) -| (t3);
                    \draw (v) |- (t4);
                    \draw (v) |- (t5);
                    \draw (v) |- (l);
                \end{tikzpicture}%}}}
            }
            \caption{Constructing a tree from its subtrees.}
            \label{FIG:tree}
        \end{minipage}
        \hfil%
        \begin{minipage}{.32\textwidth}
            \centering
            \subcaptionbox{Loop\label{SUBFIG:loop}}{%
                \begin{tikzpicture}[scale=.3]%{{{
                    \tikzset{every path/.style={thick,rounded corners=1mm}}
                    \tikzset{every node/.style={full}}

                    \node (v) at (1,2) {};
                    \node (a) at (3,1) {};
                    \node (b) at (2,3) {};

                    \draw (v) -| (a);
                    \draw (b) -| (v);
                    \draw (v) |- (4,0) |- (0,4) |- (v);
                \end{tikzpicture}%}}}
            } \\
            \subcaptionbox{Parallel Edges\label{SUBFIG:parallel}}[3cm]{
                \begin{tikzpicture}[scale=.2]%{{{
                    \tikzset{every path/.style={thick,rounded corners=1mm}}
                    \tikzset{every node/.style={full}}

                    \node (u)  at (2,3) {};
                    \node (v)  at (7,7) {};
                    \node (uS) at (1,1) {};
                    \node (uE) at (4,2) {};
                    \node (uN) at (3,4) {};
                    \node (vW) at (5,8) {};
                    \node (vS) at (6,5) {};
                    \node (vE) at (8,6) {};

                    \draw (v)  |- (-1,9) |- (u);
                    \draw (v)  |- (9,10) |- (0,0) |- (u);
                    \draw (v)  |- (vS);
                    \draw (vE) |- (v);
                    \draw (vW) |- (v);
                    \draw (u)  |- (uS);
                    \draw (u)  |- (uN);
                    \draw (uE) |- (u);
                \end{tikzpicture}%}}}
            }
            \caption{Multigraphs with \complexity{} two.}
        \end{minipage}%
    \end{figure}

    \medskip
    In order to compute a confluent orthogonal representation, using dynamic
    programming, only $O(\deg(v'))$ steps are required for each vertex $v'$.
    Thus, the total time complexity is linear.
    \qed
\end{proof}

%%%%%%%%%%%%%%%%%%%%%%%%%%%%%%%%%%%%%%%%%%%%%%%
%
%%%%%%%%%%%%%%%%%%%%%%%%%%%%%%%%%%%%%%%%%%%%%%%

\section{Multi-Graphs}\label{SEC:multi}

\begin{theorem}\label{THEO:multi-examples}
    There are 4-modal multigraphs that need \complexity{} at least two in any
    \PCOD{}.
\end{theorem}
\begin{proof}
    Consider the digraph containing a loop in \cref{SUBFIG:loop} or the digraph
    containing two parallel edges in \cref{SUBFIG:parallel}.
    The incident 4-modal vertices on the one hand and the rotation of the outer
    face on the other hand, imply that the loop and one of the two parallel
    edges, respectively, must have \complexity{} two.
    In the case of the loop, the angle at the vertex is convex.
    Since the rotation of the outer face is $-4$, it follows that five concave
    bends on the loop are needed.
    In the case of two parallel edges, the angles at the vertices in the outer
    face are zero.
    Thus, the two edges together must have eight concave bends.
    Since each edge has an odd number of bends, there must be an edge with five
    bends.
    \qed
\end{proof}

\begin{figure}[t]%{{{
    \center
    \newcommand\drawingscale{.3}
    \newlength\drawingunit
    \setlength\drawingunit{.4cm}

    \begin{subfigure}{.19\textwidth}
        \center
        \vspace{2\drawingunit}
        \begin{tikzpicture}[scale=\drawingscale,baseline=\drawingscale*4cm]%{{{
            \tikzset{every path/.style={thick,rounded corners=2}}
            \coordinate (v1) at (4,4);
            \fill (v1) circle (0.2);
            \draw (v1) -- (4,0);
            \draw (v1) -- (4,6);
            \draw (v1) -| (3,6);
            \draw (v1) -| (6,6);
            \draw (v1) |- (1,1) -- (1,0);
            \draw (v1) |- (2,2) -- (2,6);
            \draw (v1) |- (5,5) -- (5,6);
            \draw (v1) |- (7,3) -- (7,6);
        \end{tikzpicture}%}}}
        \caption{$+$}
        \label{fig:heu:source-sw}
    \end{subfigure}
    \hfil%
    \begin{subfigure}{.19\textwidth}
        \center
        \vspace{2\drawingunit}
        \begin{tikzpicture}[scale=\drawingscale,baseline=\drawingscale*4cm]%{{{
            \tikzset{every path/.style={thick,rounded corners=2}}
            \coordinate (v1) at (4,4);
            \fill (v1) circle (0.2);
            \draw (v1) -- (4,1);
            \draw (v1) -- (4,6);
            \draw (v1) -| (3,6);
            \draw (v1) -| (6,6);
            \draw (v1) -| (7,1);
            \draw (v1) |- (1,2) -- (1,1);
            \draw (v1) |- (2,3) -- (2,6);
            \draw (v1) |- (5,5) -- (5,6);
        \end{tikzpicture}%}}}
        \vspace{1\drawingunit}
        \caption{$+-$}
        \label{fig:heu:oi}
    \end{subfigure}
    \hfil%
    \begin{subfigure}{.19\textwidth}
        \center
        \vspace{1\drawingunit}
        \begin{tikzpicture}[scale=\drawingscale,baseline=\drawingscale*4cm]%{{{
            \tikzset{every path/.style={thick,rounded corners=2}}
            \coordinate (v1) at (4,4);
            \fill (v1) circle (0.2);
            \draw (v1) -- (4,1);
            \draw (v1) -- (4,7);
            \draw (v1) -| (2,1);
            \draw (v1) -| (6,7);
            \draw (v1) |- (1,6) -- (1,1);
            \draw (v1) |- (3,2) -- (3,1);
            \draw (v1) |- (5,5) -- (5,7);
            \draw (v1) |- (7,3) -- (7,7);
        \end{tikzpicture}%}}}
        \vspace{1\drawingunit}
        \caption{$+-+$}
    \end{subfigure}
    \hfil%
    \begin{subfigure}{.19\textwidth}
        \center
        \vspace{1\drawingunit}
        \begin{tikzpicture}[scale=\drawingscale,baseline=\drawingscale*4cm]%{{{
            \tikzset{every path/.style={thick,rounded corners=2}}
            \coordinate (v1) at (4,4);
            \fill (v1) circle (0.2);
            \draw (v1) -- (4,2);
            \draw (v1) -- (4,7);
            \draw (v1) -| (2,2);
            \draw (v1) -| (6,7);
            \draw (v1) -| (7,2);
            \draw (v1) |- (1,6) -- (1,2);
            \draw (v1) |- (3,3) -- (3,2);
            \draw (v1) |- (5,5) -- (5,7);
        \end{tikzpicture}%}}}
        \vspace{2\drawingunit}
        \caption{$+-+-$}
    \end{subfigure}
    \hfil%
    \begin{subfigure}{.19\textwidth}
        \center
        \begin{tikzpicture}[scale=\drawingscale,baseline=\drawingscale*4cm]%{{{
            \tikzset{every path/.style={thick,rounded corners=2}}
            \coordinate (v1) at (4,4);
            \fill (v1) circle (0.2);
            \draw (v1) -- (4,2);
            \draw (v1) -- (4,8);
            \draw (v1) -| (2,2);
            \draw (v1) -| (6,2);
            \draw (v1) |- (1,7) -- (1,2);
            \draw (v1) |- (3,3) -- (3,2);
            \draw (v1) |- (5,6) -- (5,8);
            \draw (v1) |- (7,5) -- (7,2);
        \end{tikzpicture}%}}}
        \vspace{2\drawingunit}
        \caption{$+-+-+$}
    \end{subfigure}

    \bigskip
    \begin{subfigure}{.19\textwidth}
        \center
        \vspace{1\drawingunit}
        \begin{tikzpicture}[scale=\drawingscale,baseline=\drawingscale*4cm]%{{{
            \tikzset{every path/.style={thick,rounded corners=2}}
            \coordinate (v1) at (4,4);
            \fill (v1) circle (0.2);
            \draw (v1) -- (4,6);
            \draw (v1) -| (1,1);
            \draw (v1) -| (2,2) -| (8,6);
            \draw (v1) -| (3,6);
            \draw (v1) -| (6,6);
            \draw (v1) |- (5,5) -- (5,6);
            \draw (v1) |- (7,3) -- (7,6);
        \end{tikzpicture}%}}}
        \caption{$-$}
        \label{fig:heu:sink-sw}
    \end{subfigure}
    \hfil%
    \begin{subfigure}{.19\textwidth}
        \center
        \vspace{1\drawingunit}
        \begin{tikzpicture}[scale=\drawingscale,baseline=\drawingscale*4cm]%{{{
            \tikzset{every path/.style={thick,rounded corners=2}}
            \coordinate (v1) at (4,4);
            \fill (v1) circle (0.2);
            \draw (v1) -- (4,1);
            \draw (v1) -- (4,6);
            \draw (v1) -| (1,1);
            \draw (v1) -| (3,6);
            \draw (v1) -| (6,6);
            \draw (v1) |- (2,2) -- (2,1);
            \draw (v1) |- (5,5) -- (5,6);
            \draw (v1) |- (7,3) -- (7,6);
        \end{tikzpicture}%}}}
        \caption{$-+$}
        \label{fig:heu:io}
    \end{subfigure}
    \hfil%
        \begin{subfigure}{.19\textwidth}
        \center
        \vspace{1\drawingunit}
        \begin{tikzpicture}[scale=\drawingscale,baseline=\drawingscale*4cm]%{{{
            \tikzset{every path/.style={thick,rounded corners=2}}
            \coordinate (v1) at (4,4);
            \fill (v1) circle (0.2);
            \draw (v1) -- (4,2);
            \draw (v1) -- (4,6);
            \draw (v1) -| (1,2);
            \draw (v1) -| (3,6);
            \draw (v1) -| (6,6);
            \draw (v1) -| (7,2);
            \draw (v1) |- (2,3) -- (2,2);
            \draw (v1) |- (5,5) -- (5,6);
        \end{tikzpicture}%}}}
        \vspace{1\drawingunit}
        \caption{$-+-$}
        \label{fig:heu:ioi}
    \end{subfigure}
    \hfil%
    \begin{subfigure}{.19\textwidth}
        \center
        \begin{tikzpicture}[scale=\drawingscale,baseline=\drawingscale*4cm]%{{{
            \tikzset{every path/.style={thick,rounded corners=2}}
            \coordinate (v1) at (4,4);
            \fill (v1) circle (0.2);
            \draw (v1) -- (4,2);
            \draw (v1) -- (4,7);
            \draw (v1) -| (1,2);
            \draw (v1) -| (3,7);
            \draw (v1) -| (6,2);
            \draw (v1) |- (2,3) -- (2,2);
            \draw (v1) |- (5,6) -- (5,7);
            \draw (v1) |- (7,5) -- (7,2);
        \end{tikzpicture}%}}}
        \vspace{1\drawingunit}
        \caption{$-+-+$}
        \label{fig:heu:ioio}
    \end{subfigure}
    \hfil%
        \begin{subfigure}{.19\textwidth}
        \center
        \begin{tikzpicture}[scale=\drawingscale,baseline=\drawingscale*4cm]%{{{
            \tikzset{every path/.style={thick,rounded corners=2}}
            \coordinate (v1) at (4,4);
            \fill (v1) circle (0.2);
            \draw (v1) -- (4,2);
            \draw (v1) -| (2,2);
            \draw (v1) -| (5,6) -| (0,2);
            \draw (v1) -| (7,2);
            \draw (v1) |- (1,5) -- (1,2);
            \draw (v1) |- (3,3) -- (3,2);
            \draw (v1) -| (6,6) -- (6,7);
        \end{tikzpicture}%}}}
        \vspace{1\drawingunit}
        \caption{$-+-+-$}
        \label{fig:heu:ioioi}
    \end{subfigure}
    \caption{%
        Drawings around $v_k$ in the proof of \cref{THEO:multi}.
        A $+$ represents outgoing edges from below.
        A $-$ represents incoming edges from below.%
    }
    \label{fig:heu}
\end{figure}
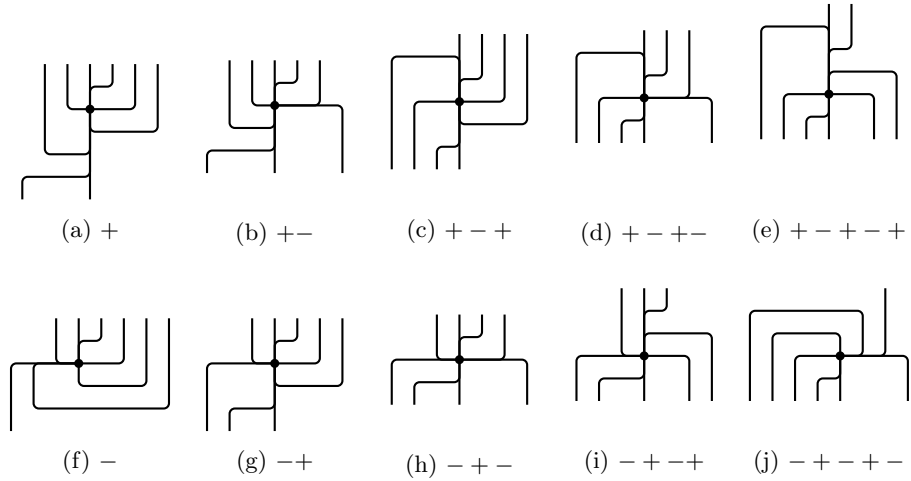%}}}

\begin{theorem}\label{THEO:multi}
    Every 4-modal multigraph has a \PCOD{} with \complexity{} at most two.
    Moreover, such a drawing can be computed in linear time.
\end{theorem}
\begin{proof}
    The approach is inspired by \cite{biedl/kant:98}.
    Subdivide each loop.
    Let the resulting digraph be $G$.
    Then make the digraph biconnected maintaining its
    4-modality~\cite{angelini_etal:jgaa22}.
    Now compute in linear time \cite{brandes:esa02} an $st$-ordering
    $v_1,\dots,v_n$ of this biconnected graph $G'$ (without taking into account
    the direction of the edges).
    Iteratively add the vertices with increasing y-coordinates in the order of
    the $st$-ordering, maintaining a column for each edge that has exactly one
    end vertex drawn.

    Let $v_k$ be a vertex.
    An edge $e$ incident to $v_k$ is incident to $v_k$ \emph{from below} if $e$
    has an end vertex that is before $v_k$ in the $st$-ordering.
    Let $e_1,\dots,e_j$ be the sequence of edges incident to $v_k$ from below as
    they appear from left to right.
    Since $v_k$ is 4-modal, $e_1,\dots,e_j$ can be divided into at most five
    subsequences of edges consisting only of incoming ($-$) or only of outgoing
    ($+$) edges of $v_k$.
    Depending on the arrangement of these subsequences, we assign the bends
    around $v_k$.
    E.g., consider the Case~$+-$ in \cref{fig:heu:oi}, i.e., among the edges
    incident to $v_k$ from below, there are first some outgoing edges, followed
    by some incoming edges.
    All outgoing edges from below are attached to the South port, while all
    incoming edges from below are attached to the East port.
    Mind that all outgoing edges except one need two bends near $v_k$.
    Consider now the edges incident to $v_k$ to later vertices in the
    $st$-ordering.
    By 4-modality, there can be at most some incoming edges, followed by some
    outgoing, some incoming and again some outgoing edges in counter-clockwise
    order around $v_k$.
    We attach them to the East, North, West, and South port of $v_k$,
    respectively.
    The edges from below determine the position of $v_k$.
    In our case, $v_k$ is drawn above one of the edges attached to its South
    port.

    See \cref{fig:heu} for the routing of the edges from below and the possible
    edges to later end vertices in the other cases.
    For $v_1$, we choose the assignment according to \cref{fig:heu:source-sw} or
    \cref{fig:heu:sink-sw}.
    After all vertices are placed, we remove edges that are in $G'$ but not in $G$.
    If $x$ is a vertex that was inserted into a loop, we reroute the two
    incident edges near $x$ such that the incoming edge of $x$ has exactly one
    bend near $x$ and the outgoing edge has no bend near~$x$.
    Finally, we eliminate zig-zags.

    By the $st$-ordering, the columns of the edges incident to $v_k$ from below
    are consecutive among the edges with exactly one end vertex drawn
    \cite{biedl/kant:98}.
    This implies planarity if the columns for the new edges are inserted
    directly next to $v_k$.
    For each edge $e$, there are at most two bends near the tail of $e$ and at
    most three bends near the head of $e$.
    Consider now a 2-cycle $(v,x),(x,v)$ replacing a loop at $v$.
    Since the subdivision vertex $x$ is incident to exactly one incoming and one
    outgoing edge, it follows that near $x$ there is no bend on $(x,v)$ and one
    bend on $(v,x)$.
    If $(x,v)$ does not have three bends near $v$ then in total there are at
    most six bends on the loop, namely the four bends near $v$ plus one bend
    near $x$ plus the bend on $x$.
    Since the number of bends on an edge must be odd, there are only five.
    Consider now the case that $(x,v)$ has three bends near $v$
    (\cref{fig:heu:sink-sw,fig:heu:ioioi}).
    If in addition $(v,x)$ has two bends near $v$, then there are seven bends on
    the loop.
    However, in this case, there is a zig-zag on $(v,x)$ formed by the bend near
    $x$ and the second bend near $v$.
    Thus, after eliminating zig-zags, the \complexity{} is at most two.
    \qed
\end{proof}

%%%%%%%%%%%%%%%%%%%%%%%%%%%%%%%%%%%%%%%%%%%%%%%
%
%%%%%%%%%%%%%%%%%%%%%%%%%%%%%%%%%%%%%%%%%%%%%%%

\section{Irreducible Triangulations}\label{SEC:irred}

%An \emph{irreducible triangulation} is an internally triangulated graph with an
%outer face of degree four that does not contain any separating triangles.
%
We prove that every 4-modal digraph whose underlying undirected graph is an
irreducible triangulation has a \PCOD{} with \complexity{} at most one.

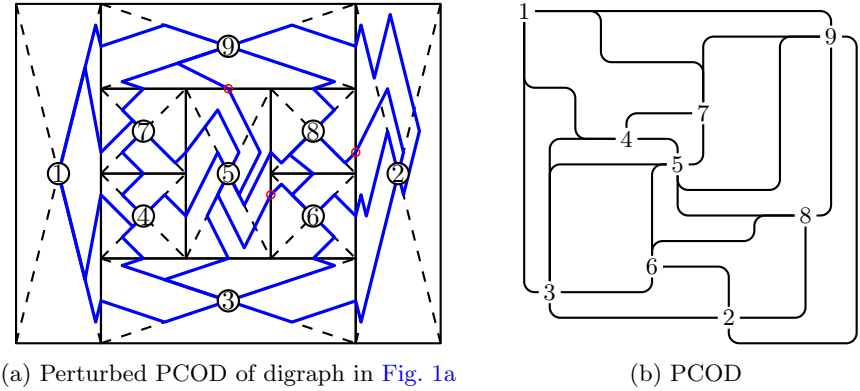
\begin{figure}[t]
    \centering
    \subcaptionbox{Perturbed PCOD of digraph in \cref{SUBFIG:graph}\label{SUBFIG:perturbed}}[0.6\textwidth]{%
        \includegraphics[page=1]{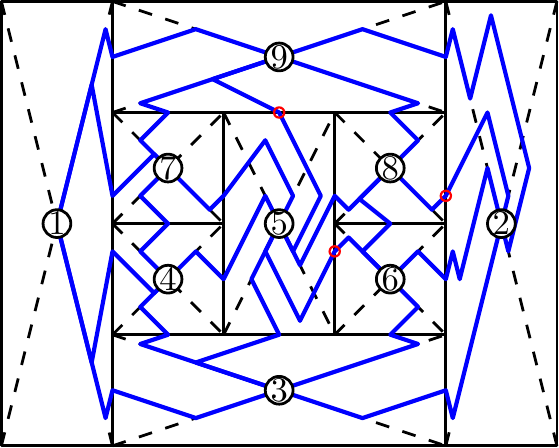}
    }
    \hfil%
    \subcaptionbox{PCOD\label{SUBFIG:pcodFromPerturbed}}{%
        \begin{tikzpicture}[scale=.34]%{{{
            \tikzset{every path/.style={thick,rounded corners}}
            \tikzset{every node/.style={circle,inner sep=0.04}}

            \node (1) at (1,14) {$1$};
            \node (2) at (9,2) {$2$};
            \node (3) at (2,3) {$3$};
            \node (4) at (5,9) {$4$};
            \node (5) at (7,8) {$5$};
            \node (6) at (6,4) {$6$};
            \node (7) at (8,10) {$7$};
            \node (8) at (12,6) {$8$};
            \node (9) at (13,13) {$9$};

            \draw (7) |- (4,12) |- (1);
            \draw (9) |- (1) ;
            \draw (1) |- (3,11) |- (4);
            \draw (1) |- (3);
            \draw (6) |- (5);
            \draw (7) |-  (9);
            \draw (7) |-  (5);
            \draw (4) |-  (7);
            \draw (3) |- (4);
            \draw (3) |- (5);
            \draw (6) |-  (3);
            \draw (3) |- (2);
            \draw (2) |- (14,1) |- (9);
            \draw (2) |- (6);
            \draw (9) |- (8);
            \draw (5) |- (11,7) |- (9);
            \draw (5) |- (4);
            \draw (5) |- (8);
            \draw (6) |- (10,5) |- (8);
            \draw (8) |- (2);
        \end{tikzpicture}%}}}
    }
    \hfil%
    \caption{%
        Perturbed \PCOD{} and corresponding \PCOD{} after zig-zag elimination.
        Red encircled bends are due to the change of the coordinate system
        and~not~real.
    }
    \label{FIG:perturbed}
\end{figure}

Motivated by the approaches in \cite{angelini_etal:jgaa22,biedl/mondal:arxiv17},
we use \emph{rectangular duals}, a contact representation of an irreducible
triangulation $G=(V,E)$ with the following properties.
The vertices $v \in V$ are represented by internally disjoint axis-parallel
rectangles $R(v)$.
Two rectangles touch if and only if the respective vertices are adjacent in $G$.
Moreover, no four rectangles representing a vertex meet at the same point and
$\bigcup_{v \in V} R(v)$ is a rectangle.
See the rectangles in \cref{SUBFIG:perturbed}.
A rectangular dual for an irreducible triangulation can be computed in linear
time~\cite{bhasker/sahni:88,biedl/derka:jgaa16,he:93,kant/he:wg93}.

Given a rectangular dual, we perturb the coordinate system such that in each
rectangle the axes correspond to the diagonals.
The \emph{perturbed x-axis} is the diagonal containing the bottommost-leftmost
point of the rectangle, the other diagonal is the \emph{perturbed y-axis}.
A \emph{perturbed orthogonal polyline} is a polyline such that in each rectangle
the segments are parallel to one of the axes.
A bend of a perturbed orthogonal polyline at the boundary of two rectangles is a
\emph{real bend} if among the two incident segments one is parallel to a
perturbed x-axis and the other parallel to a perturbed y-axis.
Bends inside a rectangle are always real.
In a \emph{perturbed \PCOD{}} each vertex $v$ is drawn at the center of $R(v)$.
An edge $(u,v)$ is a perturbed orthogonal polyline in $R(u) \cup R(v)$ between
$u$ and $v$ starting with a segment on the perturbed y-axis in $R(u)$ and ending
with a segment on the perturbed x-axis in $R(v)$.
The drawing of $(u,v)$ must have at least one bend in the interior of both
$R(u)$ and $R(v)$ and must cross the boundary of $R(u)$ and $R(v)$ exactly once.
Distinct edges may overlap in a first or last segment, but must not intersect
otherwise.
No two bends have the same coordinates.
See \cref{SUBFIG:perturbed}.
The \emph{North port} of $v$ is the port above and to the left of the center of
$R_v$.
The\emph{West, South, and East ports} are the other ports in counter-clockwise
order.

In analogy to the arguments in \cite{angelini_etal:jgaa22}, we obtain that a
perturbed \PCOD{} yields a confluent orthogonal representation where the number
$s$ of left turns counts only real bends.
By the next theorem, we can derive a \PCOD{} with \complexity{} one from a
suitable perturbed \PCOD{} after zig-zag elimination.
See~\cref{SUBFIG:pcodFromPerturbed}.

\begin{theorem}\label{THEO:irred}
    Every 4-modal irreducible triangulation has a \PCOD{} with \complexity{} at
    most one; and such a drawing can be computed~in~linear~time.
\end{theorem}
\begin{proof}
    Let $G$ be an irreducible triangulation.
    We construct a rectangular dual for $G$ ignoring edge directions.
    Routing the edges inside any rectangle independently, we then construct a
    perturbed \PCOD{} that yields a confluent orthogonal representation with
    \complexity{} at most one after zig-zag elimination.

    Let $v$ be a vertex of $G$.
    For a side $s$ of $R(v)$ let $u_i$, $i=1,\dots,k$ be the adjacent vertices
    of $v$ in counter-clockwise order such that $s$ and $R(u_i)$ intersect in
    more than a point.
    Let $e_i$ be the edge between $v$ and $u_i$.
    Consider the division of $\left<e_1,\dots,e_k\right>$ into
    \emph{mono-directed classes}, i.e., maximal subsequences such that any two
    edges in a subsequence are either both incoming or both outgoing edges.
    Let the modality mod$(s)$ of $s$ be the number of these subsequences.
    Since $G$ is 4-modal we have mod$(s) \leq 5$.
    Assume now that $s$ is a side of $R(v)$ with maximal modality.
    Assume without loss of generality that $s$ is the right side of~$R(v)$.

    \begin{figure}[t]
        \centering
        \subcaptionbox{\label{SUBFIG:mod5out}mod$(s)=5$}{%
            \includegraphics[page=2]{rectangular}
        }
        \hfil%
        \subcaptionbox{\label{SUBFIG:mod5zigzag}with zig-zag}{%
            \includegraphics[page=1]{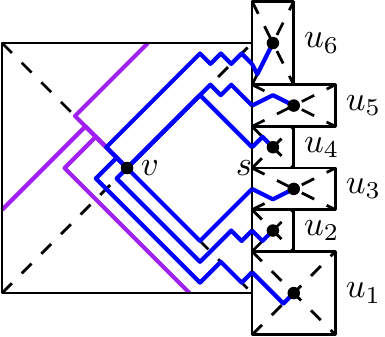}
        }
        \hfil%
        \subcaptionbox{\label{SUBFIG:mod5in}in first}{%
            \includegraphics[page=3]{rectangular}
        }
        \\[1ex]
        \subcaptionbox{\label{SUBFIG:mod4in}mod $4$}{%
            \includegraphics[page=6]{rectangular}
        }
        \hfil%
        \subcaptionbox{\label{SUBFIG:mod4outC}mod $4$}{%
            \includegraphics[page=8]{rectangular}
        }
        \hfil%
        \subcaptionbox{\label{SUBFIG:mod3in}mod $4$}{%
            \includegraphics[page=17]{rectangular}
        }
        \hfil%
        \subcaptionbox{\label{SUBFIG:mod4outCC}mod $4$}{%
            \includegraphics[page=7]{rectangular}
        }
        \hfil%
        \subcaptionbox{\label{SUBFIG:mod3in}mod $4$}{%
            \includegraphics[page=18]{rectangular}
        }
        \\[1ex]
        \subcaptionbox{\label{SUBFIG:mod3in}mod $3$}{%
            \includegraphics[page=9]{rectangular}
        }
        \hfil%
        \subcaptionbox{\label{SUBFIG:mod3out}mod $3$}{%
            \includegraphics[page=10]{rectangular}
        }
        \hfil%
        \subcaptionbox{\label{SUBFIG:mod2in}mod $2$}{%
            \includegraphics[page=11]{rectangular}
        }
        \hfil%
        \subcaptionbox{\label{SUBFIG:mod2outC}mod $2$}{%
            \includegraphics[page=13]{rectangular}
        }
        \hfil%
        \subcaptionbox{\label{SUBFIG:mod2outCC}mod $2$}{%
            \includegraphics[page=12]{rectangular}
        }

        \caption{Routing the edges in a perturbed \PCOD{}.}
        \label{FIG:rectangularAssignment}
    \end{figure}

    \paragraph{mod$(s)$ = 5.}
    If $e_1$ is an outgoing edge of $v$, assign the mono-directed classes of
    edges crossing $s$ from bottom to top in this order to (i) the North port
    bending three times to the left, (ii) to the West port bending twice to the
    left, (iii) to the South port bending once to the left, (iv) to the East
    port bending once to the right, and (v) to the North port bending twice to
    the right.
    Route the edges as indicated in blue in \cref{SUBFIG:mod5out} to $s$.
    By adding zig-zags, it is always possible to route an edge $e_i$ between $v$
    and $u_i$ in such a way that the parts of $e_i$ in $R(v)$ and $R(u_i)$ meet
    in $s$.
    See \cref{SUBFIG:mod5zigzag}.
    Edges crossing other sides of $R_v$ are all outgoing edges of $v$ and are
    assigned to the North port, bending once or twice in the direction of the
    side where they leave $R_v$.
    See the purple edges in \cref{SUBFIG:mod5out}.
    If $e_1$ is an incoming edge, start analogously with the West port.
    See \cref{SUBFIG:mod5in}.

    \paragraph{mod$(s) \in \{1,\dots,4\}$.}
    The assignment of edges to ports and the routing of the edges are contained
    in the drawing of the case mod$(s)=5$.
    See the blue edges in the second and third row in
    \cref{FIG:rectangularAssignment}.
    We make again sure that an edge to a side of $R(v)$ with modality one has at
    most two bends in the interior of $R(v)$.
    In order to do so, we have to take special care if mod$(s)=4$ and the
    bottommost edge is an outgoing edge of $v$.
    Let $s_t$ and $s_b$ be the top and bottom side, respectively, of $R(v)$.
    If $v$ is incident to an outgoing edge crossing $s_t$, we opt for the
    variant in \cref{SUBFIG:mod4outC} and otherwise (including the case that $v$
    is incident to an incoming edge crossing $s_b$) for the variant in
    \cref{SUBFIG:mod4outCC}.
    No particular care has to be taken if mod$(s)=2$
    (\cref{SUBFIG:mod2outC,SUBFIG:mod2outCC}).

    \medskip

    Let now $e$ be an edge between two vertices $u$ and $v$.
    We consider the number $b_e(u)$ and $b_e(v)$ of bends on $e$ in $R(u)$ and
    $R(v)$, respectively, after eliminating zig-zags.
    We assume without loss of generality that $b_e(u) \leq b_e(v)$.
    Recall that then $1 \leq b_e(u) \leq b_e(v) \leq 3$.
    Let $s_u$ and $s_v$ be the sides of $R(u)$ and $R(v)$, respectively that
    contain the intersection $s$ of $R(u)$ and $R(v)$.

    \begin{figure}[t]
        \centering
        \subcaptionbox{\label{SUBFIG:bendOns}real bend on $s$}{%
            \includegraphics[page=16]{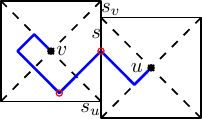}
        }
        \hfil%
        \subcaptionbox{\label{SUBFIG:manybends}$b_e(v) \geq 3$ implies \dots}{%
            \includegraphics[page=14]{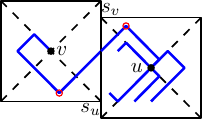}
        }
        \hfil%
        \subcaptionbox{\label{SUBFIG:bu2}\dots $b_e(u) \leq 2$}{%
            \includegraphics[page=15]{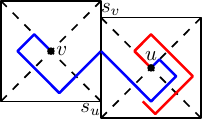}
        }
        \hfil%
        \caption{Eliminating zig-zags to reduce the number of bends per edge to
            three.}
        \label{FIG:eliminatingZigZags}
    \end{figure}

    We have to show that up to zig-zags there are in total at most three bends
    on~$e$.
    This is clear if $b_e(u)=b_e(v) = 1$.
    Assume now that $b_e(v) \geq 2$.
    Since the number of real bends on $e$ is odd it follows that the bend on $s$
    is real if and only if $b_e(u)+b_e(v)$ is even.
    In this case the bend on $s$ bends in opposite direction as the next bend in
    $R(u)$ and $R(v)$ (otherwise $e$ does not cross $s$).
    Since $b_e(v) \geq 2$, the real bend of $e$ on $s$ and the next bend of $e$
    in $R(v)$ form a zig-zag and can be eliminated.
    See \cref{SUBFIG:bendOns}.
    Thus, if $b_e(u)+b_e(v) \leq 4$ then there are at most three bends on $e$
    after zig-zag elimination.

    It remains to consider the case that $b_e(v) = 3$ and $b_e(u) \geq 2$.
    This implies mod$(s_v)>1$ and $e$ is in the first or last mono-directed
    class among the edges crossing $s_v$.
    We assume without loss of generality that $s_v$ is the right side of $R_v$
    and that $e$ is in the bottommost mono-directed class.
    See \cref{SUBFIG:manybends}.
    It follows that $e$ is an outgoing edge of $v$ and thus, an incoming edge
    of $u$.
    Since $b_e(u)\geq 2$ it follows that $e$ is attached to the East port of
    $u$.
    Assume first that $b_e(u) = 2$.
    Then the bends of $e$ in $R(u)$ are in opposite direction as the bends of
    $e$ in $R(v)$.
    Thus, there is at least one zig-zag consisting of a bend in $R(u)$ and a
    bend in $R(v)$.
    After eliminating this zig-zag there are only three bends left.

    Assume now that $b_e(u)=3$.
    This is only possible if mod$(s_u)\geq 2$.
    Hence, $R(u)$ is the topmost or bottommost rectangle incident to the right
    of $R(v)$.
    Since $e$ is in the bottommost class with respect to $s_v$, it must be the
    bottommost one.
    Thus, $R(v)$ is the topmost neighbor to the left of $R(u)$.
    Moreover, since mod$(s_u)\geq 2$ there must be a port of $u$ other than the
    East port that contains an edge $e'$ (red edge in \cref{SUBFIG:bu2})
    crossing $s_u$.
    But $e'$ would have to bend at least four times in the interior of $R(u)$,
    which never happens according to our construction.
    \qed
\end{proof}

%%%%%%%%%%%%%%%%%%%%%%%%%%%%%%%%%%%%%%%%%%%%%%%
%
%%%%%%%%%%%%%%%%%%%%%%%%%%%%%%%%%%%%%%%%%%%%%%%

\section{(Quasi-)Upward-Planar Drawings}\label{SEC:upward}

\begin{theorem}\label{THEO:upward}
    Every upward-plane digraph admits an upward \PCOD{} with \complexity{} at
    most one.
    Moreover, for plane $st$-graphs both the \complexity{} and the total number
    of bends can be minimized simultaneously in linear time.
\end{theorem}
\begin{proof}
    Let $G$ be an upward-plane digraph.
    Then $G$ can be augmented to a plane $st$-graph by adding edges
    \cite{dibattista/tamassia:88}.
    Subdividing each edge once yields a plane $st$-graph with  a bitonic
    $st$-ordering \cite{angelini_etal:wg20} and, thus, with an upward-planar
    L-drawing \cite{chaplick_etal:gd17}.
    This corresponds to an upward \PCOD{} of \complexity{} one for $G$.

    If $G$ is a plane $st$-graph it can be decided in linear time whether $G$
    has an upward-planar L-drawing \cite{chaplick_etal:gd17}, and thus, an
    upward \PCOD{} of \complexity{} zero.
    Otherwise, the minimum number of edges that has to be subdivided in order to
    obtain a digraph that has a bitonic $st$-ordering can be computed in linear
    time \cite{angelini_etal:wg20}.
    Thus, a \PCOD{} with the minimum number of bends among all upward \PCODs{}
    of $G$ with \complexity{} one can be computed in linear time.
    Observe that the total number of bends cannot be reduced by increasing the
    \complexity{}, since the subdivision of edges is only performed in order to
    break one of the transitive edges in a valley.
    \qed
\end{proof}

\begin{figure}
    \subcaptionbox{empty East}{%
        \begin{tikzpicture}[scale=0.4]%{{{
            \tikzset{every path/.style={thick,rounded corners}}
            \node (v) at (0,0) {$v$};
            \draw (-2,1) |- (v) ;
            \draw (-3,-1) |- (v) ;
            \draw (v) |- (-1,2);
            \draw (v) |- (1,3);
            \draw[dashed] (v) |- (-4,-2);
            \draw[dashed] (v) |- (2,-3);
        \end{tikzpicture}%}}}
    }
    \hfil%
    \subcaptionbox{South to North\label{SUBFIG:quasiNOEast}}[0.25\linewidth]{%
        \begin{tikzpicture}[scale=0.4]%{{{
            \tikzset{every path/.style={thick,rounded corners}}
            \node (v) at (0,0) {$v$};
            \draw (-2,1) |- (v);
            \draw (-3,-1) |- (v);
            \draw (v) |- (-1,2);
            \draw (v) |- (1,3);
            \draw[dashed] (v) |- (0.5,1.33) |- (-4,-2);
            \draw[dashed] (v) |- (1,1.66) |- (2,-3);
        \end{tikzpicture}%}}}
    }
    \hfil%
    \subcaptionbox{\mbox{East/West full}}[0.25\linewidth]{%
            \begin{tikzpicture}[scale=0.4]%{{{
                \tikzset{every path/.style={thick,rounded corners}}
                \node (v) at (0,0) {$v$};
                \draw (-2,1) |- (v) ;
                \draw (-3,-1) |- (v) ;
                \draw[dashed] (1,2) |- (v) ;
                \draw[dashed] (2,-1) |- (v) ;
                \draw[dashed] (v) |- (-4,-2);
                \draw[dashed] (v) |- (2,-3);
            \end{tikzpicture}%}}}
    }
    \hfil%
    \subcaptionbox{East to West\label{SUBFIG:quasiEastWest}}{%
        \begin{tikzpicture}[scale=0.4]%{{{
            \tikzset{every path/.style={thick,rounded corners}}
            \node (v) at (0,0) {$v$};
            \draw (-2,1) |- (v);
            \draw (-3,-1) |- (v);
            \draw[dashed] (-1.5,2) |- (v);
            \draw[dashed] (2,-1) |- (-1,1) |- (v);
            \draw[dashed] (v) |- (-4,-2);
            \draw[dashed] (v) |- (2,-3);
        \end{tikzpicture}%}}}
    }
    \caption{From a planar L-drawing of a 2-modal graph to a quasi-upward \PCOD{}.}
    \label{FIG:quasi}
\end{figure}
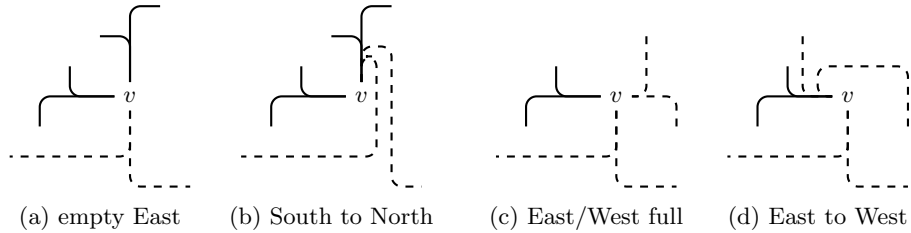

\begin{theorem}\label{THEO:quas-upward}
    Every 2-modal digraph without 2-cycles admits a quasi-upward \PCOD{} with
    \complexity{} at most one.
    Moreover, such a drawing can be computed in linear time.
\end{theorem}
\begin{proof}
    Let $G$ be a 2-modal graph without 2-cycles.
    $G$ has a planar L-drawing~\cite{angelini_etal:jgaa22}, say~$\Gamma$.
    Process the vertices $v$ of $G$ top-down in~$\Gamma$.
    If there are edges attached to the South port of $v$, we reroute them such
    that they are attached to the North port.
    Consider first the case that at least one among the East or the West
    port~--~say the East port~--~of $v$ does not contain edges.
    Then we can reroute the edges as indicated in \cref{SUBFIG:quasiNOEast}.
    Edges attached to the South port of $v$ in $\Gamma$ are now attached to the
    North port of $v$ and get two additional bends near their tail.

    Assume now that both the East and the West port of $v$ contain edges.
    Then, by 2-modality, no edge is attached to the North port of $v$.
    Those edges incident to the East port that bend upward are reattached to the
    West port without adding any additional bend.
    This is possible since these edges have already been rerouted near their
    other end vertex and two new bends have been inserted.
    So there are enough bends for the bend-or-end property.
    The rotation of the faces and the angular sum around the vertices are also
    maintained.
    The edges incident to the East port bending downward are rerouted from their
    original drawing to the West port with two new bends.
    See \cref{SUBFIG:quasiEastWest}.
    Now we can reroute the edges attached to the South port as in the first
    case.

    An edge attached to the South port in $\Gamma$ gets at most two new bends
    near its tail; an edge attached to the North port at most two new bends near
    its head.
    Thus, in the end each edge has at most three bends, i.e., \complexity{} 1.
    \qed
\end{proof}

%%%%%%%%%%%%%%%%%%%%%%%%%%%%%%%%%%%%%%%%%%%%%%%
%
%%%%%%%%%%%%%%%%%%%%%%%%%%%%%%%%%%%%%%%%%%%%%%%

\section{Experiments using an ILP}\label{SEC:ilp}

Based on the definition of confluent orthogonal representations and the fact
that each 4-modal multigraph has a \PCOD{} with \complexity{} at most two, we
developed an ILP to compute \PCODs{} with minimum \complexity{} for 4-modal
graphs.
See the appendix.
%See \cite{cornelsen/diatzko:2022} for details.
Since each simple 4-modal graph without 2-cycles can be extended to a
triangulated 4-modal graph \cite{angelini_etal:jgaa22}, we first sampled several
thousand upward-planar triangulations for various numbers $n \leq 500$ of
vertices with two different methods:
sampling (a) undirected triangulations uniformly at random
\cite{poulalhon/schaeffer:algorithmica2006} orienting the edges according to an
$st$-ordering \cite{brandes:esa02} and (b) with an OGDF method \cite{ogdf}.
Then we flipped the direction of each edge with probability 0.5 maintaining
4-modality.
Finally, we added as many 2-cycles as 4-modality allowed.
The resulting digraphs contained $(\frac{3}{4}\pm\frac{1}{4})n$ separating
triangles, roughly $n$ 2-cycles, but no separating 2-cycles.
All digraphs had split~complexity~one.

%%%%%%%%%%%%%%%%%%%%%%%%%%%%%%%%%%%%%%%%%%%%%%%
%
%%%%%%%%%%%%%%%%%%%%%%%%%%%%%%%%%%%%%%%%%%%%%%%

\section{Conclusion and Future Work}

We examined the \complexity{} of \PCODs{} of various graph classes.
In particular, we have shown that every 4-modal digraph admits a \PCOD{} with
\complexity{} two even if it contains loops and parallel edges and that
\complexity{} two is sometimes necessary.
For simple digraphs, we made a first step, by proving that every 4-modal
irreducible triangulation admits a \PCOD{} with \complexity{} one.
It still remains open whether \complexity{} one suffices for all simple 4-modal
digraphs.
Experiments suggest that this could very well be true.
It would also be interesting to know whether the minimum \complexity{} or the
minimum number of bends in a \PCOD{} or a (quasi-)upward \PCOD{} can be
efficiently determined in the case of a given 4-modal, 2-modal, or upward-planar
embedding, respectively, as well as in the case when no embedding is given.

\bibliographystyle{splncs04}
\bibliography{biblio}

\newpage

\appendix

\section*{Appendix: ILP for Minimizing \SComplexity{}}

Based on the definition of a confluent orthogonal representation and on the fact
that each 4-modal graph has a \PCOD{} with \complexity{} at most two
(\cref{THEO:multi}), we use an ILP in order to find \PCODs{} with minimum
\complexity{}.

For each entry $r$ in the confluent orthogonal representation, we use the
variables $s_r^{(1)}, s_r^{(2)}, b_r^L, b_r^R \in \{0,1\}$, $a_r \in \{0,1,2\}$,
and $e_r \in \{0,1\}$ as follows:
\begin{itemize}
    \item
        $a[r] = a_r  \cdot \pi + \begin{cases}
            0 & \textrm{if $e[r]$ and $e[\overline{r_p}]$ is a switch at $v[r]$} \\
            \frac\pi2 & \textrm{otherwise}
        \end{cases}$
    \item
        $e_r=1$ if and only if the number $s[r]$ of left turns on $e[r]$
        traversed from $v[r]$ is odd.
        The variables $s_r^{(i)}$,$s_{\overline r}^{(i)}$, $i=1,2$ represent a
        first and a second pair of bends on $e[r]$, where $s_r^{(i)}=1$ if the
        two bends are left turns on $e[r]$ traversed from $v[r]$.
        I.e., $$s[r] = 2(s_r^{(1)}+s_r^{(2)}) + e_r.$$
        Recall that by \cref{THEO:multi} there are at most five bends on any
        edge  in a \PCOD{} with minimum \complexity{}.
    \item
        $b_r^L$  and $b_r^R$ together represent $b[r]$, namely $b_r^L=1$ if and
        only if $b[r] = L$ and $b_r^R=1$ if and only if $b[r] = R$.
\end{itemize}
In order to ensure consistency, we add the following constraints:
\begin{align*}
    s_r^{(1)} + s_{\overline r}^{(1)} &\leq 1 &
    s_r^{(2)} + s_{\overline r}^{(2)} &\leq 1 &
    b_r^L + b_r^R &\leq 1
\end{align*}

\Cref{PROP:rotation} and \cref{PROP:angularSum} of feasible confluent orthogonal
representations are straightforwardly formulated as linear constraints.
\cref{PROP:coveredBends} translates to
$$s[r] - b_r^L - b_{\overline r}^R \geq 0.$$
\cref{PROP:bendorend} can be formulated as
$$a_r+b_r^R+b_{\overline{r_p}}^L \geq 1
\textup{ if } e[r] \textup{ and } e[\overline{r_p}] \textup{ is a switch at } v[r].$$
In order to fulfill \cref{PROP:oddbends}, we require
$$e_r + e_{\overline r} = 1.$$
The objective function is
\[ \min \sum_r \left( s_r^{(1)} + |E| \cdot s_r^{(2)} \right). \]
This way, the number of edges with five bends is minimized first, then that of
edges with three bends.

In a practical implementation, one might fix $s_r^{(2)} = 0$ at first to check for a
solution with \complexity{} one and only allow $s_r^{(2)} \in \{0,1\}$ if no such
solution exists.

\end{document}